\documentclass[10pt,conference]{IEEEtran}
\IEEEoverridecommandlockouts

\usepackage{cite}
\usepackage{amsmath,amssymb,amsfonts}
\usepackage{algorithmic}
\usepackage{graphicx}
\usepackage{textcomp}
\usepackage[table,xcdraw]{xcolor} 
\def\BibTeX{{\rm B\kern-.05em{\sc i\kern-.025em b}\kern-.08em
T\kern-.1667em\lower.7ex\hbox{E}\kern-.125emX}}

\usepackage{booktabs}
\usepackage{tabularx}
\usepackage{array}
\usepackage[hidelinks]{hyperref}  
\usepackage{cleveref}
\usepackage{makecell}
\usepackage{pifont}
\usepackage{soul}
\usepackage{changes}
\usepackage{comment}
\usepackage{dblfloatfix}
\usepackage{adjustbox} 
\usepackage[numbers,square,sort&compress]{natbib}

\definecolor{strongagree}{HTML}{FFFFFF}
\definecolor{agree}{HTML}{FFFFFF}
\definecolor{neither}{HTML}{FFFFFF}
\definecolor{disagree}{HTML}{FFFFFF}
\definecolor{strongdisagree}{HTML}{FFFFFF}

\definecolor{AIMBlue}{HTML}{3871E0}
\definecolor{AIMGold}{HTML}{EEAE3C}
\definecolor{QuoteText}{HTML}{284F9B}

\begin{document}

\title{Understanding Google Summer of Code: Related Works Analysis}
\title{The Increased Importance of Mentors in the era of AI: what changes, what remains, and what is accentuated}
\title{The Constant and the Shift: OSS Mentoring in the Age of AI}
\title{More Human, Not Less: OSS Mentoring in the Generative AI Era}
\title{Mentorship Accentuated: A GSoC Case Study of Mentoring in the Age of AI}
\title{Mentorship Accentuated: What AI Changes in GSoC Mentoring and What It Cannot.}
\title{Mentorship Accentuated: Where AI Disrupts the GSoC Pipeline, and Where It Cannot}
\title{Mentorship, Interrupted: How GenAI Disrupts the GSoC Pipeline and Why Mentors Matter More \\ 
}

\makeatletter
\newcommand{\authorrowbreak}{%
  \end{@IEEEauthorhalign}
  \hfill\mbox{}\par\vspace{0.8em}
  \mbox{}\hfill\begin{@IEEEauthorhalign}
}
\makeatother
\newcommand{\fixedauthorname}[1]{%
  \IEEEauthorblockN{\makebox[0.31\textwidth][c]{#1}}%
}

\author{
\IEEEauthorblockN{
    Ruizhe Xing\textsuperscript{*},
    Zixuan Feng\textsuperscript{$\ddagger$},
    Mary Radomile\textsuperscript{$\S$},
    Stephanie Taylor\textsuperscript{$\S$},
    Sophia Vargas\textsuperscript{$\S$},
    Anita Sarma\textsuperscript{*}
}
\IEEEauthorblockA{
    \textsuperscript{*}Oregon State University, Corvallis, OR, USA, \{xingru, anita.sarma\}@oregonstate.edu\\
    \textsuperscript{$\ddagger$}Virginia Commonwealth University, Richmond, VA, USA, fengz3@vcu.edu, Co-first\\
    \textsuperscript{$\S$}Google, Mountain View, CA, USA,
    \{maryr, sttaylor, sophiavargas\}@google.com\\[4pt]
}
}

\maketitle


\begin{abstract}
Generative AI (GenAI) makes open source software (OSS) contributions cheaper to produce, and it makes them say less about the people who produced them. In OSS mentoring, that matters: mentors judge what newcomers understand from their work, and newcomers learn from mentors' feedback. We studied how GenAI is changing mentoring in Google Summer of Code (GSoC), a program in which today's mentees can become tomorrow's mentors. We screened every thread on GSoC's restricted mentors' mailing list from November 2022 through March 2026 and interviewed 15 mentors and organization administrators. GenAI helped with onboarding, coordination, and first-pass review, but proposals, entry contributions, and mentees' answers became weaker evidence of what newcomers could do. Mentors made up the difference by checking, and that checking drew on the same time and goodwill the program needs to renew itself; one organization had withdrawn from GSoC. Responses ranged from banning GenAI to accepting it, and the boundary an organization drew determined who did the checking. We propose AIM, a decision framework that links acceptable GenAI use, integration into the mentoring pipeline, and mentoring capacity, to help organizations choose responses they can sustain.
\end{abstract}

\begin{IEEEkeywords}
Google Summer of Code, GSoC, Mentoring, Open Source Software,  OSS, Sustainability.
\end{IEEEkeywords}

%

\section{Introduction} \label{sec:intro}

Generative AI (GenAI)\footnote{We use \textbf{``GenAI''} and \textbf{``AI''} interchangeably to mean \textbf{generative AI} throughout this paper.} has made open source software (OSS) contributions cheaper to produce than ever, but no cheaper to review or maintain~\cite{he2026speed,yang2026governance}. Much of the discussion of this imbalance concerns reviewer workload~\cite{afroz2026slop}. A less visible cost lies in what review is for. In OSS, review is also where mentoring happens: mentors read a newcomer's work to judge what the newcomer understands, and their feedback is how the newcomer learns~\cite{feng2024guiding}. Mentoring has always relied on the work revealing something about the person who produced it. GenAI weakens that link. A pull request that compiles and passes its tests no longer tells a mentor whether its author could have written it, can explain it, or will be able to maintain it.

This matters because OSS depends on newcomers becoming its next maintainers and mentors~\cite{trainer2014community,tan2023understanding}. Formal mentoring programs make that dependence explicit. Google Summer of Code (GSoC) has paired newcomers with mentors in more than 1,000 OSS organizations since 2005~\cite{gsoc_about}, and some of its contributors later return as mentors~\cite{trainer2014community}. The program works as a cycle: today's mentees become the people who mentor tomorrow's. If mentors cannot tell what their mentees have learned, the cycle weakens at both ends. Newcomers may finish without the skills the program was meant to build, and mentors may spend their limited time checking work instead of teaching. GSoC's 2026 GenAI guidance leaves the rules to each participating organization~\cite{gsoc_ai_guidance2026}, so each organization is working out its answer on its own.

Research has begun to document GenAI's effects on OSS, but one effect at a time. Studies have measured its effects on developer productivity~\cite{peng2023impact,becker2025productivity}, code ownership~\cite{martinlopez2026ownership}, and project-level code quality~\cite{he2026speed}; characterized the GenAI policies that projects adopt~\cite{yang2026governance,hora2026policies}; and described how AI-generated submissions strain community capacity~\cite{afroz2026slop}. Others have proposed AI mentors for onboarding~\cite{tan2025aimentor} and outlined risks for OSS learning and governance~\cite{feng2026charting}. Each of these studies examines a contribution, a project, or a policy. None follows newcomers through a mentoring program, from selection to the point where they might become mentors themselves. Yet that is where GenAI's effects meet: a weak signal at selection becomes extra checking during review, and extra verification load becomes a reason a mentor does not return. Without that view, programs and organizations cannot tell which responses protect newcomers' learning and which shift the work onto mentors.

We studied how GenAI is changing mentoring in GSoC. GSoC is a revelatory case~\cite{yin2017case}: with authorized access to the program's restricted mentors' mailing list, we could read how mentors across organizations discussed GenAI as their concerns surfaced. We screened every thread from ChatGPT's release in November 2022~\cite{openai2022chatgpt} through March 2026 and analyzed those that concerned GenAI and its impacts on mentoring in GSoC. We then interviewed 15 mentors and organization administrators (Org Admins) who differed in role, mentoring experience, organization size, and view of GenAI. We asked:

\begin{itemize}
    \item \textbf{RQ1}: How does GenAI affect mentoring across GSoC's pipeline, and what does it mean for the program's ability to renew itself?
    \item \textbf{RQ2}: How do mentors and organizations respond, and what support do they expect from GSoC?
\end{itemize}

GenAI helped mentors with onboarding, coordination, and first-pass review. Its larger effect ran the other way. Proposals, entry contributions, and even mentees' answers to review questions became weaker evidence of what newcomers understood and could do. Mentors made up the difference by checking: they interviewed applicants, reviewed code without assuming its author had examined it, and enforced GenAI rules. That checking came out of the same time and goodwill on which mentoring and the program's renewal depend. Mentors described fatigue, and one organization had withdrawn from GSoC because the work exceeded its capacity. Responses ranged from banning GenAI for specific outputs to accepting it and using it in mentoring. Across that range, the boundary an organization drew determined who did the checking. Rules about authorship depended on attributing code to GenAI, which mentors found close to impossible, and left the checking with mentors. Rules about understanding let mentors check compliance by asking contributors to explain their work, which is also how mentors teach.

The question GSoC faces is shared by any OSS community that relies on mentoring to bring in its next maintainers: how to trust newcomers without exhausting the people who build that trust. For GSoC, our results also show where program support could help organizations that now make these decisions alone. This paper makes three contributions: (1) \textbf{An account of how GenAI's effects on mentoring compound across a formal program}, from selection through review to renewal, and converge on mentor capacity. (2) \textbf{An analysis of mentoring and governance responses} showing that where a project draws its GenAI boundary determines who carries the verification work.
(3) \textbf{AIM\protect\raisebox{-2pt}{\protect\includegraphics[height=10pt]{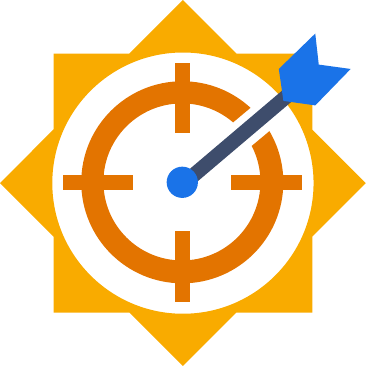}}, a mentoring and governance decision framework} that links \emph{Acceptable GenAI use}, \emph{Integration into the mentoring pipeline}, and \emph{Mentoring capacity} to help organizations choose responses they can sustain.

\section{Background: GSoC and Its Mentoring Pipeline} 
\label{sec:gsoc}


To understand GSoC and its participation patterns, mentoring structures, engagement mechanisms, and contribution outcomes, we first reviewed academic literature across four digital libraries frequently used in software engineering research \cite{feng2024guiding}: ACM Digital Library, IEEE Xplore, ScienceDirect, and SpringerLink. 
We retained only studies reporting data from GSoC participants, mentors, or projects. This ad hoc review \cite{ralph2022paving} yielded seven studies \cite{popoola2025empirical, silva2020theory, trainer2014community, raoofian2025openment, silva2017long, tan2023understanding, silva2020google}. Backward and forward snowballing on these studies for additional work on mentoring structures added three more~\cite{balali2018, steinmacher2019, feng2024guiding}. These studies characterize mentoring in OSS and give us a comprehensive vocabulary for describing associated mentoring activities. Finally, we consulted the official GSoC program documentation~\cite{gsoc_how_it_works}, which described the GSoC's timeline, roles, and evaluation points.


\begin{figure}[t!]

    \centering
    \includegraphics[width=0.9\columnwidth]{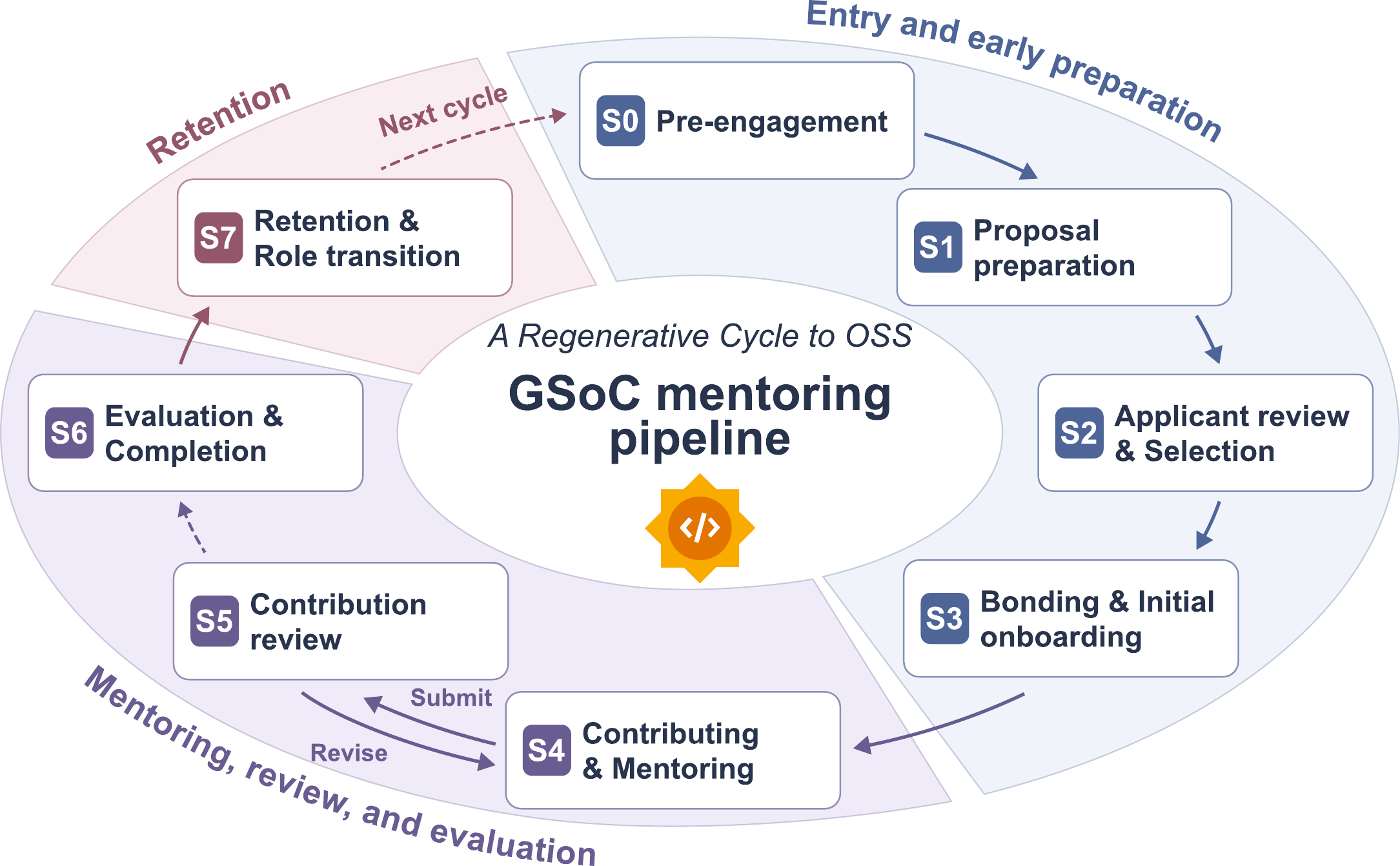}
    \caption{The literature-informed GSoC mentoring pipeline S0--S7.}
    \label{fig:mentoring-pipeline}
\end{figure}

\textbf{Google Summer of Code (GSoC)} is a global, online mentoring program run by Google that onboards newcomers to OSS. Since its launch in 2005, it has involved more than 1,000 OSS organizations~\cite{gsoc_about}. Organizations propose project ideas and select contributors based on their submitted proposals. The selected contributors, who receive stipends, then move through community bonding, project development, and evaluations~\cite{gsoc_how_it_works}. Mentors guide contributors' technical work and integration into the community, while Org Admins coordinate participation and support both mentors and contributors~\cite{gsoc_roles}. Organizations follow a shared program schedule but set their own project goals and mentoring practices~\cite{gsoc_how_it_works,gsoc_roles}. GSoC's longevity, organizational reach, and structured mentoring laid the foundation for current OSS mentoring practices. It made it a valuable case for understanding how formal OSS mentoring programs operate and the challenges they face in supporting newcomers ~\cite{silva2017long,trainer2014effects,silva2020google}.

\textbf{Literature-Informed Pipeline Development}. \Cref{fig:mentoring-pipeline} presents eight GSoC contribution stages (S0--S7), based on the official timeline~\cite{gsoc_how_it_works,gsoc_timeline} and literature on mentoring activities and roles, grouped by activity into three categories:

\textit{Entry and early preparation (S0--S3).} The GSoC begins with pre-engagement (S0), when prospective newcomers find a project of interest, contact its community, and sometimes make early contributions~\cite{silva2017long,silva2020google}. Applicants then turn that interest into a feasible project plan through proposal preparation (S1)~\cite{trainer2014community,silva2020theory}. In applicant review and selection (S2), mentors use these proposals to assess applicants' readiness and project fit and select contributors within the organization's mentoring capacity~\cite{silva2020theory,trainer2014community,tan2023understanding}. During bonding and initial onboarding (S3), selected contributors learn project tools, workflows, and community norms~\cite{balali2018,steinmacher2019,tan2023understanding}.

\textit{Mentoring, review, and evaluation (S4--S6).} As these working relationships develop, contributors move into substantive project work during contributing and mentoring (S4), building skills through mentors' guidance and feedback~\cite{silva2020theory,popoola2025empirical,feng2024guiding,tan2023understanding}. Their work moves through contribution review (S5), where review and revision inform decisions about acceptance or integration~\cite{silva2020theory,popoola2025empirical}. On GitHub, issue, pull-request, and commit interactions connect contributors and mentors in an ongoing cycle between S4 and S5~\cite{raoofian2025openment}. Alongside this cycle, evaluation and graduation (S6) assess contributors' progress and participation through milestones and midterm and final pass/fail decisions~\cite{silva2020google,tan2023understanding}.

\textit{Retention (S7).} After graduation, contributors may continue participating in their communities, sometimes returning as mentors~\cite{silva2020google,trainer2014community,tan2023understanding}. This transition renews the mentoring cycle, as former contributors draw on their experience to guide a new cohort (\Cref{fig:mentoring-pipeline}). This pipeline describes the GSoC lifecycle. In the following section, we use this as an analytic lens for investigating how GenAI affects mentoring activities and responsibilities across these stages.

\begin{figure*}[t!]

    \centering
    \includegraphics[width=\textwidth]{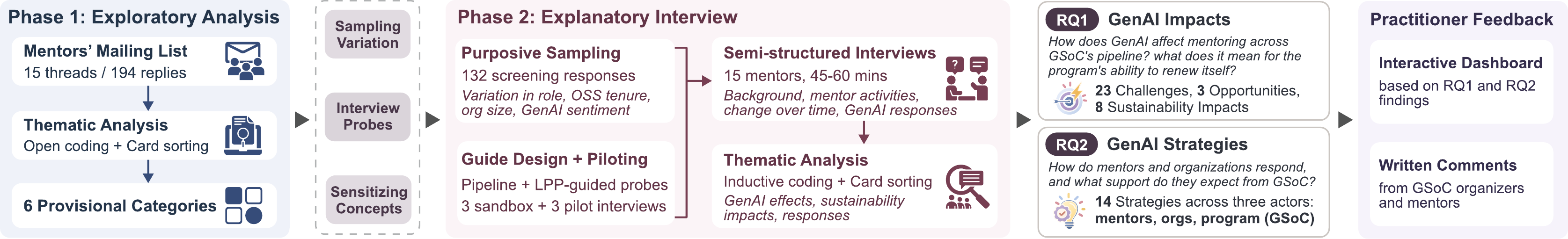}
    \caption{An overview flowchart of the research design for our qualitative, multi-method case study of GSoC.}
    \label{fig:method-flowchart}

\end{figure*}

\section{Method} \label{sec:method}

We conducted a sequential, qualitatively driven multi-method case study~\cite{runeson2009} of GSoC, which we treat as a revelatory case~\cite{yin2017case}. We first explored 
concerns and responses related to GenAI in GSoC through discussions on the mentors' mailing list. We then interviewed 15 mentors and organization administrators from GSoC to develop an explanatory understanding of their experiences, rationales, and contexts. \Cref{fig:method-flowchart} summarizes our study design.

\subsection{Phase 1: Exploratory Analysis}


\textbf{Mailing-List Screening and Corpus Construction.} 
We examined the restricted-access GSoC mentors' mailing-list archive from November 2022, the month of ChatGPT's public release~\cite{openai2022chatgpt}, through March 2026. 
Two authors independently screened every discussion thread for content broadly related to GenAI and its impacts on mentoring. They compared selections in weekly meetings and resolved disagreements through negotiated agreement~\cite{campbell2013agreement}. The final corpus comprised 15 threads and 194 replies dated between March 2025 and March 2026, involving mentors affiliated with 25 OSS organizations.

\textbf{Open Coding and Theme Development.} We conducted thematic analysis using open coding~\cite{corbin2014basics} and card sorting~\cite{zimmermann2016}. Two authors independently open-coded each reply within each thread from our final corpus, compared their interpretations in weekly meetings, and refined code definitions through discussion. They then used card sorting to group related codes and develop themes capturing patterns in GenAI-related impacts and responses in GSoC mentoring. In total, we identified six provisional themes: \textit{GenAI impacts on GSoC sustainability}, \textit{GenAI manifestations in GSoC}, \textit{Stance for GenAI use}, \textit{GenAI detection and verification}, \textit{Response and mitigation}, and \textit{Potential governance and policy gaps}.

\subsection{Phase 2: Explanatory Interviews} 

Building on Phase 1's provisional themes, we conducted semi-structured interviews~\cite{kallio2016} with 15 GSoC mentors. 

\textbf{Interview Design.} Three sources informed the interview design:
(1) LPP describes how newcomers learn through involvement in a community's practices and progression toward fuller participation~\cite{lave1991situated}. In the context of GSoC, this perspective guided questions about access to project work and knowledge, collaboration with mentors, and progression toward core contributor roles.
(2) The mentoring pipeline situated these questions within specific mentoring activities across the GSoC lifecycle.
(3) The mailing-list findings informed probes on mentoring interactions, mentor effort, reasons behind stances on GenAI, responses to its use, and potential governance gaps.

\textbf{Interview Protocol.} Our interview began with demographic and background questions about participants' OSS and mentoring experience, GSoC involvement, project context, and GenAI use in OSS development. We then asked about their roles, motivations for joining GSoC, its perceived contributions to themselves, their projects, and mentees. Derived from the pipeline, we set interview probes around mentor activities such as onboarding, mentoring interactions, contribution review, knowledge sharing, and recognition of mentoring work. We revisited these activities through comparisons with earlier cycles, asking, for example, whether onboarding practices remained effective or changed and how mentor responsibilities and effort had shifted with GenAI. Questions about GenAI responses cover those enacted and planned by mentors and organizations and their desired program support. The interview concluded with an invitation to share experiences or perspectives not covered by the preceding questions.

\textbf{Sandbox and Pilot Study.}
After developing the interview, we conducted sandbox studies with 3 software engineering researchers and revised its structure, follow-up questions, and explanations based on their feedback. For example, we placed questions about general GSoC experiences before GenAI-specific comparisons to establish participants' mentoring context and provide a reference point for discussing perceived GenAI-related changes. We then conducted 3 pilot interviews with OSS developers and refined the guide until no further suggestions emerged. These interviews prompted us to separate questions about mentoring activities, effort, and perceived effectiveness and add follow-up questions. The pilot participants were excluded from the final analysis.

\textbf{Sampling and Data Collection.}
We used stratified purposive sampling with a maximum-variation objective~\cite{campbell2020}. Before recruitment, we distributed a screening questionnaire through the GSoC mentors' mailing list between May 27 and June 9, 2026. To recruit participants with recent GSoC experience, we restricted eligibility to individuals who had served as a GSoC mentor, organization administrator, or both in at least one cycle between 2024 and 2026. The screening questionnaire further asked about role, participating cycles, OSS mentoring tenure, organization size, project categories, and GenAI sentiment. Sentiment used a five-point agreement scale with the statement ``Overall, I feel positive about the use of AI tools in OSS development.'' 

We received 132 valid responses and selected 20 candidates using a matrix of role, mentoring tenure, organization size, and GenAI sentiment. Participating cycles and project categories provided further variation. When candidates did not respond or were ineligible, we replaced them with candidates who matched as many of the original dimensions as possible. The final sample consisted of 15 participants (P1--P15) and covered every category in four sampling dimensions (\Cref{tab:demographics}). 

We conducted the interviews from June 16 to July 17, 2026. Each lasted about 45--60 minutes. Our Institutional Review Board approved this study, and participants gave informed consent to participate and be recorded. Upon completion, we offered a US\$250 donation to each participant OSS affiliation. 

\begin{table}[!h]
\centering
\caption{A demographic summary of interview participants.}
\label{tab:demographics}

\begingroup
\fontsize{9}{10}\selectfont
\setlength{\tabcolsep}{2pt}
\renewcommand{\arraystretch}{1.0}
\setlength{\extrarowheight}{0.25pt}
\setlength{\aboverulesep}{0pt}
\setlength{\belowrulesep}{0pt}
\newcommand{\demographicsheader}[1]{%
  \begin{tabular}[t]{@{}c@{}}#1\end{tabular}%
}

\begin{tabularx}{\columnwidth}{
  >{\hsize=0.45\hsize\linewidth=\hsize\centering\arraybackslash}X
  >{\hsize=0.80\hsize\linewidth=\hsize\centering\arraybackslash}X
  >{\hsize=1.00\hsize\linewidth=\hsize\centering\arraybackslash}X
  >{\hsize=1.10\hsize\linewidth=\hsize\centering\arraybackslash}X
  >{\hsize=1.65\hsize\linewidth=\hsize\centering\arraybackslash}X
}
\toprule[0.8pt]
\addlinespace[2pt]
\demographicsheader{\textbf{ID}} &
\demographicsheader{\textbf{GSoC}\\\textbf{Role}}\textsuperscript{*} &
\demographicsheader{\textbf{Mentor}\\\textbf{Tenure}\\\textbf{(years)}} &
\demographicsheader{\textbf{Org}\\\textbf{Size}}\textsuperscript{\textdagger} &
\demographicsheader{\textbf{GenAI}\\\textbf{Sentiment}\\\textbf{(screening)}} \\
\addlinespace[2pt]
\midrule

\rowcolor{black!6}
P1  & Both & 1--2 & Large
    & Strongly Agree \\
P2  & Mentor & 3--5 & Large
    & Agree \\
\rowcolor{black!6}
P3  & Both & 6--10 & Large
    & Agree \\
P4  & Mentor & 3--5 & Small
    & Strongly Disagree \\
\rowcolor{black!6}
P5  & Mentor & \textless1 & Very large
    & Disagree \\
P6  & Mentor & \textgreater10 & Very large
    & Neither \\
\rowcolor{black!6}
P7  & Both & 3--5 & Small
    & Strongly Disagree \\
P8  & Both & \textgreater10 & Very large
    & Neither \\
\rowcolor{black!6}
P9  & Admin & \textgreater10 & Small
    & Neither \\
P10 & Admin & 6--10 & Medium
    & Neither \\
\rowcolor{black!6}
P11 & Mentor & 1--2 & Large
    & Strongly Agree \\
P12 & Mentor & 1--2 & Very large
    & Disagree \\
\rowcolor{black!6}
P13 & Both & \textless1 & Small
    & Neither \\
P14 & Both & \textgreater10 & Very large
    & Neither \\
\rowcolor{black!6}
P15 & Mentor & \textgreater1 & Medium
    & Strongly Agree \\
\bottomrule[0.8pt]
\end{tabularx}

\par\vspace{2pt}
\begin{minipage}{\columnwidth}
\raggedright
\fontsize{7.5}{8.5}\selectfont
\begin{list}{}{%
  \setlength{\leftmargin}{1em}%
  \setlength{\labelwidth}{0.65em}%
  \setlength{\labelsep}{0.35em}%
  \setlength{\itemsep}{1pt}%
  \setlength{\parsep}{0pt}%
  \setlength{\topsep}{0pt}%
  \setlength{\partopsep}{0pt}%
}
\item[*] ``Both'' = GSoC mentor and Org Admin; We collectively refer to interview participants as \textbf{mentors} because both GSoC mentors and Org Admins share mentoring responsibilities in GSoC~\cite{gsoc_roles}.
\item[\textdagger] Regular contributors: small \textless10, medium 10--50, large 51--200, very large \textgreater200.
\end{list}
\end{minipage}
\endgroup
\end{table}

\textbf{Interview Analysis.} We conducted an inductive thematic analysis of the interview data as data collection progressed~\cite{braun2019, elo2008}. Two authors independently coded the transcripts and met weekly to compare interpretations and refine the codes through negotiated agreement~\cite{campbell2013agreement}. They then used card sorting~\cite{zimmermann2016} to group related codes into categories and examined the associated excerpts to refine category definitions, boundaries, and relationships. We observed indications of category saturation~\cite{francis2009} at P12. To validate the findings, we conducted three additional interviews (P13--P15), which added examples and qualifications without introducing new categories.



The analysis produced three linked codebooks: 23 GenAI \textbf{effects} on GSoC mentoring (20 challenges and 3 opportunities) and 8 \textbf{impacts} on GSoC sustainability for RQ1 (\Cref{fig:rq1}), and 14 \textbf{strategies} that participants reported for addressing these challenges for RQ2 (\Cref{tab:rq2-strategies}). 





\textbf{Practitioner Feedback.} To assess our findings and identify potential omissions, we shared an interactive dashboard of all study findings~\cite{suppl-web} available to GSoC mentors. GSoC organizers circulated its links with a feedback form asking for overall reactions, perceived strengths, points of confusion, and suggested improvements. Mentors who responded were generally favorable.





\subsection{Limitations}
\label{sec:validity}


Our study draws on GSoC through both mailing-list discussions and interviews, so the findings reflect experiences within it and may transfer partially to other formal mentoring programs in OSS. Other communities may differ in size, learning goals, project risks, and mentoring capacity, and GenAI tools and practices continue to change. Applying our findings outside GSoC requires accounting for both.

The restricted-access mailing list, where our analysis began, may make strong views or challenging experiences more visible. We deepened this analysis through explanatory interviews with 15 mentors from different backgrounds. The sample may not capture the full range of GSoC mentoring experience. Prior studies identified basic elements of overarching themes within six interviews~\cite{guest2006many} and reported code saturation at nine~\cite{hennink2017code}. We began to observe saturation at P12, and three further interviews yielded no new findings. Because we did not recruit mentees, our findings about their GenAI use, learning, and engagement reflect mentors' observations and interpretations.
\colorlet{ResultQuoteText}{QuoteText}
\newcommand{\resultsquote}[1]{\textcolor{ResultQuoteText}{\emph{``#1''}}}

\begin{figure*}[!t]

    \centering
    \includegraphics[width=0.90\textwidth]{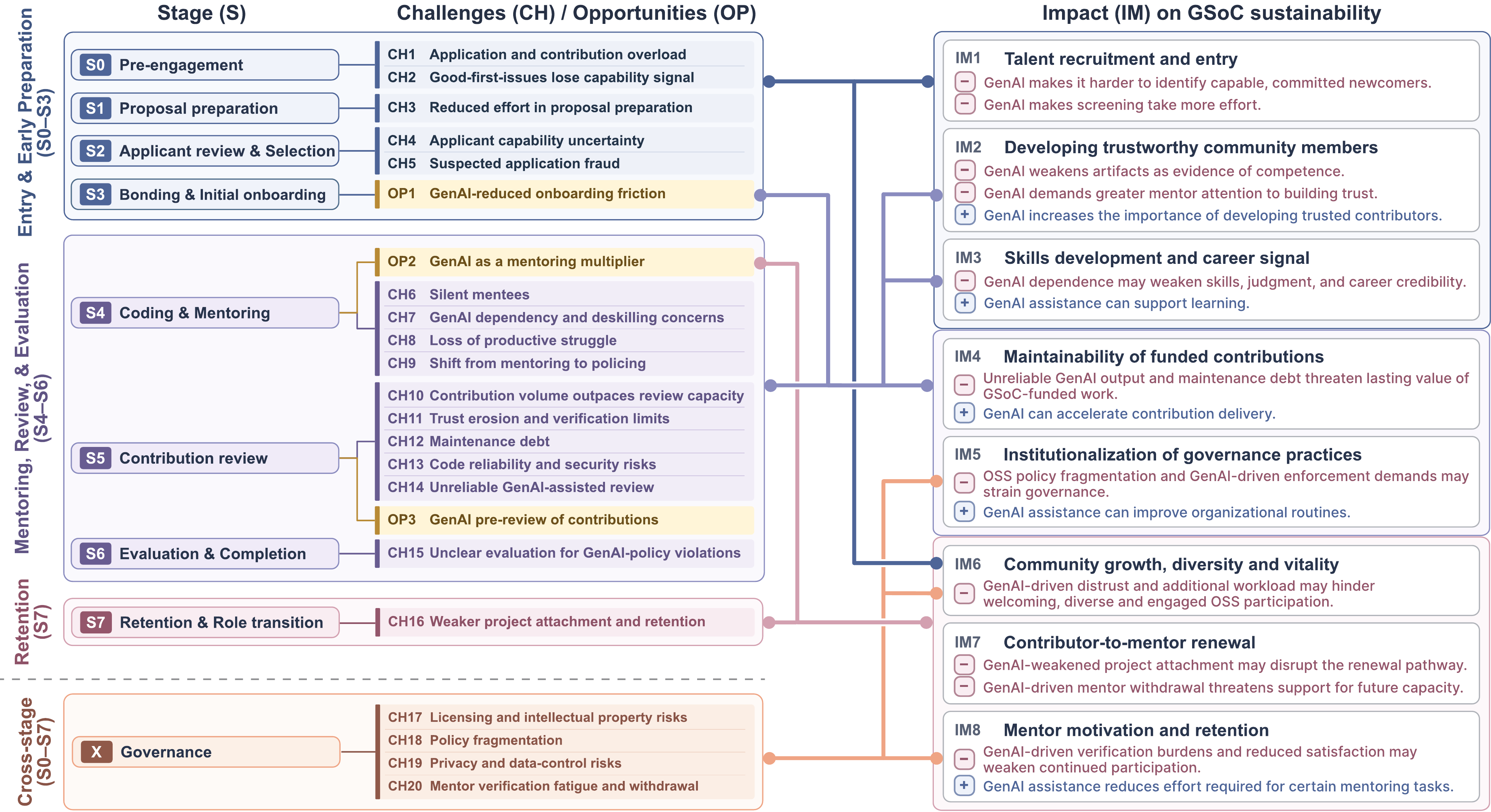}
    \caption{GenAI challenges (CH) and opportunities (OP) across the GSoC mentoring pipeline (\Cref{fig:mentoring-pipeline}) and their impacts (IM) on GSoC sustainability functions. Colored lines link groups of effects to the relevant functions. \protect\raisebox{-0.3ex}{\protect\includegraphics[height=1.8ex]{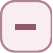}} and \protect\raisebox{-0.3ex}{\protect\includegraphics[height=1.8ex]{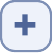}} indicate adverse and beneficial implications, respectively.}
    \label{fig:rq1}
\end{figure*}

\section{Results} \label{sec:results}


\subsection{RQ1: GenAI Impacts on GSoC} \label{subsec:result-RQ1}

\definecolor{RQOneSynthesis}{HTML}{4E6598}
\newsavebox{\rqonetakeawaybox}
\newcommand{\rqonetakeaway}[2]{%
  \par\smallskip\noindent\begingroup
  \setlength{\fboxsep}{7pt}\setlength{\fboxrule}{0.6pt}%
  \sbox{\rqonetakeawaybox}{%
    \colorbox{AIMBlue!7!white}{%
      \parbox{\dimexpr\linewidth-2\fboxsep-2\fboxrule\relax}{%
        \small{\textcolor{AIMBlue!50!black}{\textbf{Takeaway.}}} #2}}}%
  \usebox{\rqonetakeawaybox}%
  \rlap{\kern-\wd\rqonetakeawaybox\textcolor{#1}{%
    \rule[-\dp\rqonetakeawaybox]{2pt}{%
      \dimexpr\ht\rqonetakeawaybox+\dp\rqonetakeawaybox\relax}}}%
  \endgroup\par\smallskip
}


\Cref{fig:rq1} presents the full set of GenAI effects participants reported (20 challenges, CH1--CH20, and 3 opportunities, OP1--OP3), their locations in the pipeline, and their impacts on 8 GSoC sustainability functions (IM1--IM8). 
We discuss selected findings that show how these effects accumulate, organized around three concerns: \textbf{Newcomers became harder to assess and to develop} (IM1--IM3), \textbf{Mentoring time shifted toward verification} (IM4, IM5), and \textbf{Verification work reduced mentors’ motivation and strained renewal} (IM6--IM8). They draw on experiences across entry and early preparation (S0--S3), mentoring, review, and evaluation (S4--S6), retention (S7), and cross-stage governance (X). A single challenge or opportunity can affect several functions, and effects from different stages can converge on the same function.

In brief, GenAI weakened the evidence mentors use to judge newcomers and their contributions. Mentors had to establish that evidence by other means, which turned part of mentoring into verification. This work fell on the mentors whose continued participation renews the pipeline.

\textbf{Newcomers became harder to assess and to develop.} At selection (S0--S2), mentors reported that GenAI weakened entry contributions and proposals as evidence of what an applicant can do and how committed they are. 
Completed good-first-issues no longer demonstrated capability (CH2). As P15 put it, \resultsquote{They just AI it up, complete it, and say yeah, I understood everything. Even though they probably didn't.} Proposals reflected less effort (CH3); P5 received proposals that still contained the generator's reply, \resultsquote{here is the proposal you asked for}. The resulting uncertainty about applicants' capability (CH4) had concrete consequences. P4 selected an applicant whose \resultsquote{proposal looked correct} and later found that the person \resultsquote{does not know how to use Git [and] how to run shell commands}. Recruitment (IM1) thus required more effort. P8's organization had used tasks of varying difficulty to distinguish applicants, but \resultsquote{AI is meaning that those tasks are less of a differentiator and we need to meet these people in person more often, which takes more time}.

The same problem persisted once mentees began their projects (S3--S4). Participants read the change in mentor-mentee interaction differently: P15 considered GenAI useful for asking more nuanced questions, whereas silent mentees (CH6) left P14 with less insight into their reasoning. P14 recalled that earlier, \resultsquote{review went much better because you had an understanding of what the person was thinking. Now you have to be more guarded}. Asking directly did not always help. When P1 asked contributors why they had made a change, \resultsquote{They gave us an AI-generated response}.

Mentors were also concerned that GenAI displaced the learning these interactions are meant to support (IM3). Mentees could bypass the debugging through which they develop judgment (CH8). P7 described debugging with a contributor as \resultsquote{serious mental effort... [that] helped me grow along with my contributor}. In contrast, a mentee told P1, \resultsquote{I won't be able to create more PRs until my AI limit renews} (CH7). GenAI did support learning in some cases (OP1). P2 used it to prepare introductory material and \resultsquote{was able to quickly debug and answer their questions}.

Where artifacts no longer served as evidence, mentors relied on contribution history. P3 explained that with prior contributions, \resultsquote{even if [they] does an AI-generated application, we know [they are] capable}. Developing trustworthy community members (IM2) thus became both harder and more important: artifacts showed less, and a record of sustained participation remained evidence that mentors could rely on.

\textbf{Mentoring time shifted toward verification.} Because artifacts no longer served as reliable evidence, mentors had to verify both contributors' code and their compliance with GenAI rules. This verification took time from other mentoring work. \resultsquote{If it's easier to write a submission, then it's harder to review a submission...} [P15]. \resultsquote{The only significant change has been the amount of effort we're putting into reviewing PRs} [P4]. Mentors described their effort shifting from mentoring to policing (CH9). P6 had to \resultsquote{pay more attention to reminding people what the program rules were, aside from what our project-specific guidelines are}. P14 said this came at the expense of guiding mentees: \resultsquote{it's a drain on their time... [Organizations] want new people to join and grow, not... prompt a tool [and transfer review work].}

Verification began with the code, at contribution review (S5). Mentors could no longer assume that contributors had examined what they submitted (CH11): \resultsquote{I can no longer trust that a human has actually checked what they are doing} [P13]. Review thus had to establish correctness without relying on the author, while contribution volume exceeded review capacity (CH10). P1 described: \resultsquote{The code PRs are so long that it's 5,000 lines of code. And they expect us to review the complete PR within 2 days}. Accepted code could still reduce the lasting value of funded work (IM4). P8 described \resultsquote{lots of incremental and small bits and modules [that were] very costly to maintain} (CH12), and P2 warned that \resultsquote{you never know what AI missed; it could be a security threat} (CH13).

Enforcing GenAI rules added a second kind of verification. Rules varied across organizations (CH18); P4 described policies that ranged from bans to disclosure requirements to acceptance based on contribution quality. Each policy required enforcement by mentors. P12 said of disclosure, \resultsquote{[it] causes extra work for mentors... we have to stay up to date [on] whether they're following the rules and look for cues [but] that's necessary}. At evaluation (S6), mentors were also unsure how violations should affect completion decisions (CH15): \resultsquote{Every other week I see a thread about whether I should fail this contributor because they did not follow my LLM guidelines} [P12]. These enforcement demands made governance practices harder to institutionalize (IM5).

GenAI relieved parts of this work. AI pre-review (OP3) gave contributors faster feedback, since \resultsquote{the contributor can immediately get feedback... without waiting for a human} [P5], although P2 found that it caught syntax problems and missed domain logic (CH14). GenAI also eased coordination (OP2). P1 built a dashboard for unanswered issues, while P11 used AI-generated transcripts and PR summaries, noting, \resultsquote{If I had mentored three or four years ago, I would have struggled to manage the time}. These tools eased coordination and surface-level checks, but not the work of assessing contributor understanding and code quality.

\begin{table*}[!htb]

\centering
\caption{Reported and proposed strategies as responses to GenAI impacts in GSoC.}
\label{tab:rq2-strategies}
\begingroup
\fontsize{7.5}{8.5}\selectfont
\setlength{\tabcolsep}{2pt}
\renewcommand{\arraystretch}{1.0}
\setlength{\extrarowheight}{0pt}
\setlength{\aboverulesep}{1pt}
\setlength{\belowrulesep}{1pt}
\begin{tabularx}{\textwidth}{
    >{\raggedright\arraybackslash}p{2.6em}
    >{\raggedright\arraybackslash}p{0.27\textwidth}
    >{\raggedright\arraybackslash}X
}
\toprule
\textbf{ID} & \textbf{Strategy} & \textbf{Definition} \\
\midrule

\multicolumn{3}{l}{\textbf{\strut Establishing trust in applicants}} \\
\addlinespace[0.5pt]

\rowcolor{black!6}
ST1 & Selecting applicants by prior contribution trajectory & Judge applicants by months of visible work before they apply (e.g., merged PRs, follow-through on issues, and responses to review feedback) beyond proposal quality. \\
\addlinespace[1pt]

ST2 & Live interviews to verify applicants' unaided ability & Test applicants in real time through screen-share, live coding, scenario questions, or verbal explanation to confirm that they, not GenAI, can do the work. \\
\addlinespace[1pt]

\rowcolor{black!6}
ST3 & Entry tasks that require project-specific understanding & Redesign, stagger, or curate starter tasks so they depend on the project's own processes and cannot be solved by a bare prompt. \\
\addlinespace[1pt]

ST4 & Rewarding honest, rule-following GenAI use in selection & Lower the priority of applicants with undisclosed or rule-breaking GenAI use while offering revision opportunities to borderline cases. \\

\midrule[0.6pt]
\multicolumn{3}{l}{\textbf{\strut Governing GenAI use and maintaining accountability}} \\
\addlinespace[0.5pt]

\rowcolor{black!6}
ST5 & Banning GenAI for specific outputs with explicit exceptions & Restrict or ban GenAI assistance in specified outputs (e.g., proposals) or activities (e.g., onboarding) while explicitly allowing bounded uses (e.g., language help). \\
\addlinespace[1pt]

ST6 & Allowing GenAI conditionally: ``assistant, not owner'' & Permit GenAI assistance within local rules when contributors understand, check, and take ownership of the output. \\
\addlinespace[1pt]

\rowcolor{black!6}
ST7 & Judging quality, not authorship & Judge whether contributions work and meet quality expectations instead of treating who wrote them as decisive. \\
\addlinespace[1pt]

ST8 & Using GenAI for mentoring support & Use GenAI to assist routine mentoring tasks, such as onboarding, submission triage, summaries, progress tracking, or first-pass review. \\
\addlinespace[1pt]

\rowcolor{black!6}
ST9 & Teaching contributors to use GenAI responsibly & Teach contributors, through explicit instruction or by modeling practice, to read, test, and take responsibility for GenAI-assisted work while retaining human decisions. \\
\addlinespace[1pt]

ST10 & Keeping final decisions with human reviewers & Keep responsibility for reviewing and accepting submitted contributions with mentors instead of delegating verification to GenAI. \\
\addlinespace[1pt]

\rowcolor{black!6}
ST11 & Treating GenAI-suspicion cues as prompts for closer review & Use suspicious patterns of generated code or provenance checks to decide what to examine more closely, not to reject automatically, as attribution remains unclear. \\
\addlinespace[1pt]

ST12 & Asking contributors to explain their code in review & Ask contributors to explain their decisions and demonstrate their work; inability to explain is what disqualifies a contribution. \\
\addlinespace[1pt]

\rowcolor{black!6}
ST13 & Requiring contributors to disclose GenAI use & Ask contributors to disclose whether and how they used GenAI so mentors can adjust review checks targetedly. \\

\midrule[0.6pt]
\multicolumn{3}{l}{\textbf{\strut Sharing guidance and verification across the program}} \\
\addlinespace[0.5pt]
\rowcolor{black!6}
ST14 & Coordinating GenAI guidance across the program & Share guidance, resources, disclosure mechanisms, and lessons learned across participating organizations, while each organization retains autonomy over its own GenAI rules. \\

\bottomrule
\end{tabularx}
\endgroup

\end{table*}

\textbf{Verification work reduced mentors' motivation and strained renewal.} Mentors connected the accumulated verification and enforcement work to their motivation to continue (IM8). P10 raised the same concern for mentors more broadly: \resultsquote{I've been a lot more worried on the impact it has on mentors. It does create fatigue... implication in terms of workload}. In the most severe case (CH20), P7's organization withdrew from GSoC because GenAI-related demands exceeded its capacity: \resultsquote{It is just not in our bandwidth}.

Mentor withdrawal affects retention and role transition (S7), the stage where contributors become the next mentors (IM7). P3 described this renewal: \resultsquote{we are growing mentors who can guide the next generation of contributors}. Participants worried that GenAI also weakened the contributor side of this transition (CH16). P13 said, \resultsquote{I am worried [about] how much [contributors] are learning and [whether they will] stick to the project and the community}, and P11 observed that \resultsquote{folks drop off because they get automated replies or generic responses}. Renewal is therefore under pressure from both directions: contributors may not stay long enough to become mentors, and current mentors have less capacity and willingness to continue. Together with heavier screening at entry, these pressures put community growth, diversity, and vitality (IM6) at risk.

\rqonetakeaway{RQOneSynthesis}{Mentors are absorbing GenAI's costs one proposal and one pull request at a time, out of their own time and goodwill. When a mentor stops returning, GSoC loses that mentor and the future mentors that they would have guided.}

\subsection{RQ2: Mentoring and Governance Responses to GenAI} \label{subsec:result-RQ2}

Participants used or proposed 14 strategies (ST1--ST14), which respond to the three parts of RQ1. They establish trust in newcomers when artifacts no longer suffice, set how much GenAI use a project accepts while keeping contributors accountable, and seek program support for work that currently falls on mentors and organizations. Across these responses, participants' positions ranged from restricting GenAI to accepting it and using it in their own mentoring work. \Cref{tab:rq2-strategies} lists all 14, grouped by the three themes below; we discuss the strategies that carry each theme.

\textbf{Organizations established trust in applicants through new forms of evidence.} At selection, some organizations sought evidence that GenAI could not easily produce. Some relied on contribution history (ST1). P4 described qualified candidates as having \resultsquote{very good PRs already merged... interactions genuine enough to make us trust... they were reviewing their own code.} Others observed applicants at work (ST2). P3 used screen-sharing interviews to \resultsquote{make them write the code in front of us in the editor}, and P1 planned interviews to check whether applicants were \resultsquote{really strong enough without AI}. P8 proposed entry tasks that depend on project-specific understanding (ST3), \resultsquote{so that just a prompt with the task will not be able to produce the output}.

Other organizations treated how applicants used GenAI as evidence in itself (ST4). Some deprioritized applicants who used GenAI against the rules (P6, P12). Others gave applicants a chance to revise non-compliant proposals (P9, P11); P9's organization accepted applicants who \resultsquote{came back and rewrote it} after being called out. This strategy treats honesty about GenAI use as a sign of trustworthiness. It also turns a violation into an occasion for correction, so that an applicant's first misstep does not end their entry into the community.

\textbf{Responses ranged from restricting GenAI to accepting it, and all kept contributors accountable.} 
Some organizations banned GenAI for specified outputs with explicit exceptions (ST5); P5 allowed generated documentation while restricting code. Others allowed GenAI under conditions (ST6). P2 permitted it \resultsquote{as long as you know what you are working on and how it is working}, and P1 put the condition as \resultsquote{You are using AI as your assistant. AI is not your owner}. Some mentors judged contributions on quality regardless of how they were produced (ST7). P3 focused on working code, and P15 proposed prioritizing quality over flagging GenAI use. Mentors also used GenAI themselves (ST8). P4 used GitHub Copilot on larger changes because it is \resultsquote{much better than us at catching things [like] security, subtle bugs, edge cases,} and P2 followed an AI pass with a personal review: \resultsquote{one was I quickly got AI to review the whole PR code base... the second thing, I myself reviewed the PR.} P2's organization also encouraged GenAI use through written guidance: \resultsquote{We gave [contributors] written instructions on what we are expecting with the use of AI, and we encourage them to use it... [but] they shouldn't be completely dependent upon it.}

Across this range, participants held contributors accountable for what they submitted. P8 taught mentees to verify their own contributions (ST9), with the mentor \resultsquote{not being your checker}, and P15 expected that \resultsquote{if it breaks, you need to know how to identify and fix the issue}. Mentors retained final acceptance decisions (ST10); as P5 said, \resultsquote{Even after AI said it was good to go, then we have to manually review it.}

Where a project drew its boundary determined how compliance could be checked and who checked it. Rules defined by authorship require mentors to establish whether GenAI produced a contribution, and participants doubted that this was possible. P7 noted that \resultsquote{it is next to impossible to confidently attribute something [as] AI-generated,} so they \resultsquote{need to look for heuristics} and banned users only \resultsquote{when [they] do find definitive proof.} Suspicion cues, such as emoji use, outdated dependencies, and oversized PRs (ST11), could direct closer review but could not establish authorship. Rules defined by understanding pose a question that mentors can answer by asking contributors to explain and demonstrate their work (ST12). When code appeared GenAI-assisted, P8 did not accuse the mentee. They asked \resultsquote{Why did you choose to do it this way} and \resultsquote{How did you check it?}, and treated the inability to answer as a signal. P3 looked for whether a contributor \resultsquote{understands the logic and makes you understand that logic}. Under such rules, checking compliance and mentoring can take place in the same conversation. Every boundary still carried costs for mentors. Disclosure (ST13) was meant to direct review, but after GenAI disclosure, P4 \resultsquote{ran an extra set of tests} and estimated \resultsquote{maybe 20, 30\% more} review time.

\textbf{Responsibility remained local, and participants asked the program to share it.} All strategies besides ST14 are carried out by mentors and organizations and vary across organizations, consistent with the policy fragmentation reported in RQ1. Program-level support (ST14) existed only as proposals. Participants asked GSoC for two kinds of help. The first was shared guidance that would leave local choices intact. P2 called for a general guideline while noting that \resultsquote{each and every organization is differently structured. They follow their own rules along with the GSoC rules... generalized practices could be spread along.} The second was help with enforcement. P12 proposed that GSoC \resultsquote{require a strict form where the mentee discloses in what capacity they're using AI}, which \resultsquote{eases the burden from each community individually enforcing those rules}. P8 proposed program tools that would flag proposals for closer human review, \resultsquote{not to say that we would immediately disregard that person}. 


\rqonetakeaway{RQOneSynthesis}{Each organization was working out its own answer to GenAI, largely alone. The answers that asked newcomers to understand their work let mentors check it by teaching, the part of mentoring our participants valued.}


\begin{table*}[tp]
\centering
\caption[AIM: Proposed GenAI Mentoring and Governance Decision Framework across the GSoC lifecycle.]{%
  AIM\protect\raisebox{-2pt}{\protect\includegraphics[height=15pt]{figures/AIM-icon.pdf}}\enspace: Proposed GenAI Mentoring and Governance Decision Framework across the GSoC lifecycle.}
\label{tab:genai-gsoc-stances}
\begingroup
\definecolor{GenAIInk}{HTML}{233449}
\definecolor{GenAIHeaderInk}{HTML}{1F2937}
\definecolor{GenAIMuted}{HTML}{596476}
\definecolor{GenAIStageBand}{HTML}{ECEEF2}
\definecolor{GenAILabel}{HTML}{F3F4F6}
\definecolor{GenAIProh}{HTML}{92C5DE}
\definecolor{GenAISkep}{HTML}{D1E5F0}
\definecolor{GenAIAccom}{HTML}{F7F7F7}
\definecolor{GenAIPrag}{HTML}{FDDBC7}
\definecolor{GenAIAdv}{HTML}{F4A582}
\definecolor{GenAIProhTint}{HTML}{E4F0F7}
\definecolor{GenAISkepTint}{HTML}{F4F8FB}
\definecolor{GenAIAccomTint}{HTML}{FDFDFD}
\definecolor{GenAIPragTint}{HTML}{FEF6F1}
\definecolor{GenAIAdvTint}{HTML}{FCE8E0}
\definecolor{GenAIOverlay}{HTML}{E8E4F0}
\definecolor{GenAIOverlayTint}{HTML}{F8F6FB}

\newcommand{\GenAITableFont}{\fontsize{7}{7.5}\selectfont}
\newcommand{\AIMInitial}[1]{{\fontsize{10}{10}\selectfont #1}}
\newcommand{\Action}[1]{%
  \par\noindent\hangindent=0.73em\hangafter=1
  \makebox[0.73em][l]{\textbullet}#1\par\vspace{0.2pt}}
\newcommand{\StanceHeader}[3]{%
  \cellcolor{#1}\centering\arraybackslash
  \parbox[c][18pt][c]{\linewidth}{\centering\color{#2}%
    \fontsize{8.5}{10}\selectfont\bfseries #3}}
\newcommand{\LifecycleLabel}[2]{%
  \centering
  \rotatebox{90}{%
    \if\relax\detokenize{#2}\relax
      \textbf{\shortstack[c]{#1}}%
    \else
      \begin{tabular}{@{}c@{}}\textbf{#1}\\[2pt]
        {\color{GenAIMuted}#2}\end{tabular}%
    \fi}}
\newlength{\GenAICellPadding}
\newcommand{\GenAICell}[1]{%
  \raisebox{\ht\strutbox}{\vtop{%
    \hrule height0pt
    \kern\GenAICellPadding
    \hbox{\begin{minipage}[t]{\linewidth}%
      \raggedright
      #1\par\unskip
    \end{minipage}}%
    \kern\GenAICellPadding}}}

\makeatletter
\newcommand{\GenAISpectrum}{%
  \begingroup
  \dimen0=\dimexpr\linewidth*7200/7227\relax
  \edef\GenAIGradientWidthBP{\strip@pt\dimen0}%
  \immediate\pdfobj{<<
    /ShadingType 2 /ColorSpace /DeviceRGB
    /Coords [0 0 \GenAIGradientWidthBP\space 0] /Extend [true true]
    /Function << /FunctionType 3 /Domain [0 1]
      /Functions [
        << /FunctionType 2 /Domain [0 1]
           /C0 [0.572549 0.772549 0.870588]
           /C1 [0.819608 0.898039 0.941176] /N 1 >>
        << /FunctionType 2 /Domain [0 1]
           /C0 [0.819608 0.898039 0.941176]
           /C1 [0.968627 0.968627 0.968627] /N 1 >>
        << /FunctionType 2 /Domain [0 1]
           /C0 [0.968627 0.968627 0.968627]
           /C1 [0.992157 0.858824 0.780392] /N 1 >>
        << /FunctionType 2 /Domain [0 1]
           /C0 [0.992157 0.858824 0.780392]
           /C1 [0.956863 0.647059 0.509804] /N 1 >>
      ] /Bounds [0.25 0.5 0.75] /Encode [0 1 0 1 0 1 0 1]
    >>
  >>}%
  \edef\GenAIGradientObject{\the\pdflastobj}%
  \setbox0=\hbox to\linewidth{%
    \pdfliteral{q /GenAIGradient sh Q}%
    \vrule width0pt height2.4pt depth0pt\hfil}%
  \immediate\pdfxform resources{%
    /Shading << /GenAIGradient \GenAIGradientObject\space 0 R >>}0%
  \leavevmode\pdfrefxform\pdflastxform
  \endgroup}
\makeatother

\normalfont\GenAITableFont\color{GenAIInk}
\setlength{\GenAICellPadding}{8pt}
\setlength{\tabcolsep}{2pt}
\setlength{\arrayrulewidth}{1.2pt}
\renewcommand{\arraystretch}{1.0}
\arrayrulecolor{white}
\setlength{\parskip}{0pt}
\emergencystretch=1em
\renewcommand{\tabularxcolumn}[1]{p{#1}}

\newlength{\GenAILabelWidth}
\setlength{\GenAILabelWidth}{22pt}
\newlength{\GenAILabelBandWidth}
\setlength{\GenAILabelBandWidth}{\dimexpr\GenAILabelWidth+2\tabcolsep\relax}
\newlength{\GenAIGridWidth}
\setlength{\GenAIGridWidth}{\dimexpr\linewidth-\GenAILabelBandWidth-\arrayrulewidth\relax}
\newsavebox{\GenAIRowBox}
\newlength{\GenAIRowHeight}

\newcommand{\GenAIStanceRow}[3]{%
  \begingroup
  \sbox{\GenAIRowBox}{%
    \begin{minipage}[c]{\GenAIGridWidth}%
    \begin{tabularx}{\linewidth}{*{5}{>{\raggedright\arraybackslash}X|}}%
    #3\\
    \end{tabularx}%
    \end{minipage}}%
  \setlength{\GenAIRowHeight}{\dimexpr\ht\GenAIRowBox+\dp\GenAIRowBox\relax}%
  \setlength{\fboxsep}{0pt}%
  \noindent\makebox[\linewidth][l]{%
    \colorbox{GenAILabel}{%
      \parbox[c][\GenAIRowHeight][c]{\GenAILabelBandWidth}{\LifecycleLabel{#1}{#2}}}%
    \hspace{\arrayrulewidth}\usebox{\GenAIRowBox}}%
  \par\nointerlineskip\vspace{\arrayrulewidth}%
  \endgroup}

\noindent\hspace*{\dimexpr\GenAILabelBandWidth+\arrayrulewidth\relax}%
\begin{minipage}{\GenAIGridWidth}
\noindent\makebox[\linewidth]{%
  \rlap{\color{GenAIMuted}More Restrictive}%
  \hfill\textbf{\AIMInitial{A}cceptable GenAI use Stance}\hfill
  \llap{\color{GenAIMuted}More Permissive}}\par
\vspace{1pt}\noindent\GenAISpectrum
\end{minipage}\par\vspace{3pt}

\noindent\hspace*{\dimexpr\GenAILabelBandWidth+\arrayrulewidth\relax}%
\begin{minipage}{\GenAIGridWidth}
\begin{tabularx}{\linewidth}{*{5}{>{\raggedright\arraybackslash}X|}}
\StanceHeader{GenAIProh}{GenAIHeaderInk}{Prohibitionist} &
\StanceHeader{GenAISkep}{GenAIHeaderInk}{Skeptic} &
\StanceHeader{GenAIAccom}{GenAIHeaderInk}{Accommodator} &
\StanceHeader{GenAIPrag}{GenAIHeaderInk}{Pragmatist} &
\StanceHeader{GenAIAdv}{GenAIHeaderInk}{Advocate} \\
\end{tabularx}
\end{minipage}\par\nointerlineskip\vspace{\arrayrulewidth}
\GenAIStanceRow{GenAI boundary}{Stance Definition}{%
\cellcolor{GenAIProhTint}\GenAICell{%

\Action{Prohibit GenAI in specified outputs and stages.}
\Action{List permitted exceptions, such as accessibility or translation.}
\Action{Accept unpolished work rather than allowing unlisted GenAI polishing.}
}
 &
\cellcolor{GenAISkepTint}\GenAICell{%

\Action{Restrict GenAI by default, allowing only a short list of low-risk uses (e.g., code explanations).}
\Action{Require contributors to develop the substance of their work.}
\Action{Require approval and output checks for other uses.}
}
 &
\cellcolor{GenAIAccomTint}\GenAICell{%
\Action{Permit GenAI for supporting tasks (e.g., documentation or testing) rather than core work.}
\Action{List exclusions for each output or stage.}
\Action{Require contributors to review, adapt, and check output and to keep their reasoning and decisions.}
}
 &
\cellcolor{GenAIPragTint}\GenAICell{
\Action{Permit GenAI across development, including implementation; state only necessary project limits.}
\Action{Require contributors to understand and check outputs.}
\Action{Require contributors to make decisions and answer review questions themselves.}
}
 &
\cellcolor{GenAIAdvTint}\GenAICell{
\Action{Integrate GenAI into daily development within security, privacy, and licensing rules.}
\Action{Expect contributors to use GenAI as a working tool.}
\Action{Hold contributors responsible for diagnosing and fixing failures in what they submit.}
}
}

\begingroup\setlength{\fboxsep}{0pt}
\noindent\colorbox{GenAIStageBand}{%
  \parbox[c][14pt][c]{\linewidth}{\centering
  \textbf{\AIMInitial{I}ntegration into the Mentoring Pipeline}}} 
\par\nointerlineskip\vspace{\arrayrulewidth}
\endgroup
\GenAIStanceRow{Entry and early\\preparation}{}{%
\cellcolor{GenAIProhTint}\GenAICell{%
\Action{Require work covered by the ban to be done without GenAI.}
\Action{Ask applicants to demonstrate their skills directly if their readiness is unclear.}
\Action{Apply published consequences for policy violations and allow revisions where permitted.}
}
 &
\cellcolor{GenAISkepTint}\GenAICell{%
\Action{Ask applicants to disclose any permitted GenAI use and how they checked its output.}
\Action{Probe their own reasoning with follow-up questions, and confirm that any use beyond the list was approved.}
\Action{Look closely at questionable submissions through discussion before deciding.}
}
 &
\cellcolor{GenAIAccomTint}\GenAICell{%
\Action{Allow GenAI for supporting tasks in the application, but not for core proposal or entry tasks.}
\Action{Ask applicants to explain their independently developed their ideas or task solutions.}
\Action{Ask applicants to explain how they reviewed, adapted, and checked generated outputs.}
}
 &
\cellcolor{GenAIPragTint}\GenAICell{%
\Action{Accept GenAI-assisted work and assess whether it works and meets project requirements.}
\Action{Ask applicants to explain their choices, demonstrate their checks, and respond to feedback themselves.}
}
 &
\cellcolor{GenAIAdvTint}\GenAICell{%
\Action{Ask applicants to demonstrate how they use GenAI on a project-specific task, including how they provide context and evaluate suggestions.}
\Action{Ask them to demonstrate how they would diagnose and correct failures in the generated output.}
}
}
\GenAIStanceRow{Mentoring, review,\\and evaluation}{}{%
\cellcolor{GenAIProhTint}\GenAICell{%
\Action{Teach contributors to break down problems, use project documentation, and develop their own solutions for work covered by the ban.}
\Action{Ask contributors to walk through their implementation and how they tested it, and use gaps in understanding to guide further mentoring.}
\Action{Have contributors make requested revisions themselves with GenAI allowed only for listed exceptions.}
}
 &
\cellcolor{GenAISkepTint}\GenAICell{%
\Action{Teach contributors to develop core solutions themselves with GenAI use only for permitted low-risk assistance.}
\Action{Show contributors how to check GenAI suggestions against documentation, code behavior, or tests before relying on it.}
\Action{Ask contributors to explain how bounded GenAI use shaped their reasoning, and target extra review or tests at the affected parts.}
}
 &
\cellcolor{GenAIAccomTint}\GenAICell{%
\Action{Teach contributors to develop the core work themselves and use GenAI for supporting tasks.}
\Action{Ask contributors to inspect GenAI output for wrong assumptions, missing cases, and inconsistencies before mentor review.}
\Action{Ask contributors to explain core decisions and check that supporting material alongside the major changes reflects what the code does.}
}
 &
\cellcolor{GenAIPragTint}\GenAICell{%
\Action{Teach contributors to define problems, break down tasks, give GenAI useful context, and evaluate assisted or generated solutions.}
\Action{Guide contributors to adapt generated code, test it within the project, and resolve failures before submission.}
\Action{Review contributions against project requirements, and have contributors diagnose and fix work that fails, regardless of how it was produced.}
}
 &
\cellcolor{GenAIAdvTint}\GenAICell{%
\Action{Model how to give GenAI project context, inspect its outputs, and decide which to use, change, or reject.}
\Action{Bring GenAI-assisted attempts into routine mentor discussions, with contributors presenting their checks, decisions, and open questions.}
\Action{Use planned checkpoints to confirm that contributors still understand their work and can diagnose and fix failures as GenAI use grows.}
}
}
\GenAIStanceRow{Governance}{}{%
\cellcolor{GenAIProhTint}\GenAICell{
\Action{Review disputed cases to separate confirmed violations from uncertain attribution.}
\Action{Check whether the rules or the process for fixing or disputing a violation were unclear, and whether enforcement took time away from mentoring support.}
\Action{Update the examples, exceptions, and evidence expectations upon review.}
}
 &
\cellcolor{GenAISkepTint}\GenAICell{
\Action{For each permitted use, check whether contributors follow the rules and can stand behind the output.}
\Action{Keep, narrow, or expand the permitted list, expanding only when the results are consistently good.}
\Action{Document how additions are approved and checked, and keep only the disclosure that helps review.}}
 &
\cellcolor{GenAIAccomTint}\GenAICell{
\Action{Check supporting tasks for recurring inaccuracies or mismatches with the core implementation.}
\Action{Identify where GenAI help crossed into core work, obscured decisions, or shifted corrections onto mentors.}
\Action{Update the exclusions for each output or stage, and the checking instructions for supporting tasks.}
}
 &
\cellcolor{GenAIPragTint}\GenAICell{
\Action{Review GenAI-assisted work for whether contributors could explain, fix, and maintain it.}
\Action{Note where mentors had to reconstruct the reasoning or fix the work themselves.}
\Action{Revise the project limits, contributor responsibilities, and mentoring guidance where gaps appeared.}
}
 &
\cellcolor{GenAIAdvTint}\GenAICell{
\Action{Review where routine GenAI use supported learning and where contributors struggled to catch or fix faulty output.}
\Action{Update approved tools and their access limits, mentoring routines, and checkpoints to close those gaps.}
\Action{Widen project scope or automation only if mentor can still verify the work and contributors keep learning.}
}
}
\par\vspace{3pt}
\begingroup\setlength{\fboxsep}{5pt}
\colorbox{GenAIOverlay}{\parbox{\dimexpr\linewidth-2\fboxsep\relax}{%
  \centering\color{GenAIInk}\textbf{\AIMInitial{M}entoring Capacity}}}
\endgroup\par\nointerlineskip
\begin{tabularx}{\linewidth}{*{4}{>{\raggedright\arraybackslash}X|}}
\cellcolor{GenAIOverlayTint}\GenAICell{
\textbf{GenAI Boundary}\par\vspace{3pt}
\Action{Check that mentors have enough time and expertise to apply the chosen stance.}
\Action{Assign mentors to learning support and final review, and Org Admins to policy and workload coordination.}
\Action{Pause new GenAI permissions when capacity is limited.}
}
 &
\cellcolor{GenAIOverlayTint}\GenAICell{
\textbf{\shortstack[l]{Entry and early preparation}}\par\vspace{3pt}
\Action{Match contributor numbers and project scope to available mentoring capacity.}
\Action{Publish entry requirements, check them before detailed review, and stagger starter tasks when volume is high.}
\Action{Use additional screening when applicants' readiness is hard to trace.}
}
 &
\cellcolor{GenAIOverlayTint}\GenAICell{
\textbf{\shortstack[l]{Mentoring, review, and evaluation}}\par\vspace{3pt}
\Action{Budget mentor time for the work your stance adds. 
}
\Action{Reuse reviewed guides and automated checks for routine work; use GenAI support only where the stance permits.}
\Action{Have contributors check their work and prepare explanations before review.}
}
 &
\cellcolor{GenAIOverlayTint}\GenAICell{
\textbf{Governance}\par\vspace{3pt}
\Action{Check review backlogs, missed mentoring discussions, and mentor feedback at agreed intervals.}
\Action{Cap open submissions per contributor or narrow tasks when review falls behind.}
\Action{If overload persists, 
adjust estimated contributor numbers next cycle.}
}
 \\
\end{tabularx}
\arrayrulecolor{black}
\endgroup
\end{table*}

\section{Discussion} \label{sec:discussion}

Our results show how GenAI's costs to mentoring travel through the GSoC pipeline, from selection to mentors' return (\Cref{subsec:result-RQ1}), and how the GenAI boundary an organization draws decides who ends up doing the checking (\Cref{subsec:result-RQ2}). In this section, we first consider what these findings mean for GenAI governance in OSS more broadly. We then present AIM\protect\raisebox{-2pt}{\protect\includegraphics[height=10pt]{figures/AIM-icon.pdf}}, a framework intended to help organizations make these decisions while keeping their mentors' capacity in view.


\subsection{Implications for OSS GenAI Governance}

Our participants found authorship hard, and sometimes impossible, to verify (\Cref{subsec:result-RQ2}). This matters beyond GSoC, because most OSS GenAI policies are built on authorship. They typically permit GenAI-assisted contributions on the condition of disclosure or human oversight~\cite{hora2026policies}, and the governance strategies proposed for OSS center on accountability and verification~\cite{yang2026governance}.
Such policies assume that someone has the time and means to check compliance, but our results suggest that the volunteer mentors who must do this checking often have neither. We argue that rules defined by understanding fit mentoring better. When a mentor asks a contributor to explain and demonstrate their work (ST12), the same conversation checks compliance and helps the contributor learn.

Rewarding honest disclosure at selection (ST4) points to a way out of the sustainability trap~\cite{afroz2026slop}, in which restrictions that protect maintainers' time also narrow newcomer participation.
Organizations that took this approach still held newcomers responsible for how they used GenAI, but they gave applicants who had broken the rules a chance to revise their proposals. In a program meant to develop newcomers, an applicant's first misstep with GenAI can then become something to learn from instead of a reason for rejection.

\subsection{AIM\protect\raisebox{-2pt}{\protect\includegraphics[height=10pt]{figures/AIM-icon.pdf}}: A Decision Framework for Mentoring with GenAI}

We propose AIM\protect\raisebox{-2pt}{\protect\includegraphics[height=10pt]{figures/AIM-icon.pdf}} (\Cref{tab:genai-gsoc-stances}) to support these decisions. AIM\protect\raisebox{-2pt}{\protect\includegraphics[height=10pt]{figures/AIM-icon.pdf}} asks three questions of a project: what GenAI assistance it permits (\emph{Acceptable GenAI use}), how it carries that boundary into each stage of the mentoring pipeline (\emph{Integration into the mentoring pipeline}), and how much of that work its mentors can sustain (\emph{Mentoring capacity}). We developed AIM\protect\raisebox{-2pt}{\protect\includegraphics[height=10pt]{figures/AIM-icon.pdf}} from GSoC alone, although it may also be useful for other formal mentoring programs.

\textbf{Acceptable GenAI use.} Participants drew this boundary in very different places. P9 ruled out GenAI (\resultsquote{you may not use AI or LLMs in any way}), whereas P15 argued that \resultsquote{projects have to be scoped in a way that acknowledges [AI]}. We organized these positions into five stances, ordered from most to least restrictive. Under a \emph{Prohibitionist} stance, specified outputs must be human-authored, as in P7's project: \resultsquote{I have a strict ban on artificial intelligence use}. A \emph{Skeptic} stance restricts GenAI by default and expands its use only when results justify it; An \emph{Accommodator} stance permits named support tasks, such as those P12 described: \resultsquote{[mentees may] check grammar, bugs, and style... We allow [that], and that's fine}. A \emph{Pragmatist} stance permits broader use, provided contributors understand and check what GenAI produces, or as P2 said, \resultsquote{you control the AI; you don't let the AI control you}. Finally, an \emph{Advocate} stance treats GenAI as a routine part of the workflow. This stance goes beyond what participants reported doing, so it has the least empirical support; we include it to complete the range. What all five stances share is that contributors remain accountable for what they submit.


The stances differ in more than how much GenAI they permit (\Cref{tab:genai-gsoc-stances}). Consider what each one tests at entry. A restrictive project wants to know what an applicant can do without GenAI, whereas a project at the Advocate end wants to know how well the applicant uses it, for example whether they give it enough project context and notice when it is wrong. The same contrast carries into mentoring. Restrictive stances teach newcomers to break problems down and build solutions themselves; permissive stances teach them to direct GenAI, judge its output, and repair what fails. Mentors' time goes to different places as well. At the restrictive end, much of it goes to close support and to disputes over suspected violations, while toward the permissive end it goes to reviewing a larger volume of assisted work and to regular checkpoints. A project's stance thus reflects what it wants newcomers to learn, and it should also reflect where its mentors can afford to spend their time.

Under stances defined by understanding, compliance is checked by asking for an explanation, and what that explanation shows depends on when it is asked. At selection, it shows what an applicant can do independently, and organizations can combine it with contribution histories, live tasks, or project-specific entry tasks (ST1--ST3). During the program, it shows whether a contributor understands and can stand behind assisted work (ST12), and the mentor's follow-up questions become teaching. P10 described the approach this way: \resultsquote{We sometimes ask them to clarify their prompt. If they use AI right, we probably won't see it; if they use it to do their job instead of themselves for them, we will. We don’t have a philosophical rejection of AI.}

\textbf{Integration into the mentoring pipeline.} A stance becomes practice through what a project does at each stage (\Cref{tab:genai-gsoc-stances}). During entry and early preparation, the stance sets what evidence of readiness to collect: a Prohibitionist project has applicants demonstrate their skills without GenAI, while an Advocate project asks them to show how they use it on a project-specific task. Stating these expectations early helps contributors prepare; P7, for example, explained their organization's ban in a call before work began. During mentoring, review, and evaluation, the stance shapes what mentors teach and how they check understanding. Across stances, AIM\protect\raisebox{-2pt}{\protect\includegraphics[height=10pt]{figures/AIM-icon.pdf}} favors checks that double as mentoring, such as having contributors walk through their implementation and using the gaps that surface to decide what to teach next. 
At the governance stage, AIM\protect\raisebox{-2pt}{\protect\includegraphics[height=10pt]{figures/AIM-icon.pdf}} asks the project to look back at how its boundary worked, including where enforcement took time from mentoring or mentors had to reconstruct a contributor's reasoning, and to narrow or widen it for the next cycle.

\textbf{Mentoring capacity.} Whatever its stance, a project depends on mentors having time to carry it out. Restrictive and permissive approaches both create work. Restrictions require mentors to communicate boundaries and handle suspected breaches, while permission requires them to assess understanding and validate assisted work (\Cref{subsec:result-RQ1}). AIM\protect\raisebox{-2pt}{\protect\includegraphics[height=10pt]{figures/AIM-icon.pdf}} asks organizations to name who will perform each check and to estimate how much of this work their mentors can absorb. It also treats capacity as something to protect, because mentors who are overloaded in one cycle may not return for the next. A project might watch for warning signs, such as growing review queues or mentoring conversations that are repeatedly missed, and respond by narrowing its scope or taking fewer contributors. In one case, P9's organization sat out a cycle entirely because \resultsquote{we didn't have enough mentors on hand to do the work}.

\textbf{Using AIM\protect\raisebox{-2pt}{\protect\includegraphics[height=10pt]{figures/AIM-icon.pdf}} in practice.} AIM\protect\raisebox{-2pt}{\protect\includegraphics[height=10pt]{figures/AIM-icon.pdf}} can support concrete work at several points in a GSoC cycle. Before the cycle, an organization can turn its stance into published application requirements and tasks; 
an Advocate project, for instance, might ask applicants to show how they use GenAI on a small project task. While the cycle runs, the mentoring row can act as a shared checklist that tells contributors what to check and be ready to explain before they ask for review, and tells mentors what to ask. The capacity band adds signals to watch, such as review backlogs and missed mentoring discussions. Once the cycle ends, the governance row gives the project a structure for a short retrospective on whether to narrow or widen its boundary. GSoC could also use the five stances as a shared vocabulary across its organizations. Organizations could state their stance when they apply, so that contributors know what to expect before they choose a project, and a program-level disclosure form of the kind P12 proposed (\Cref{subsec:result-RQ2}) could ask for the information each stance needs.

\subsection{Empirical Grounding and Future Work}

This study draws on the program as a whole. GSoC mentors share one mailing list, and we screened every thread in its archive from ChatGPT's release in November 2022 through March 2026. Because we also recruited interviewees through this list, every mentor on it had the opportunity to take part. The 15 mentors we interviewed varied in role, mentoring experience, organization size, project type, and view of GenAI, and GSoC organizers circulated our findings to mentors for feedback (\Cref{sec:method}). AIM\protect\raisebox{-2pt}{\protect\includegraphics[height=10pt]{figures/AIM-icon.pdf}} is thus built on the experiences of mentors from organizations across GSoC.

The timing of the study also worked in our favor. We began the analysis in early January 2026, when mentors were voicing GenAI concerns on the list, and conducted interviews in June and July 2026, during the cycle. Participants could therefore tell us about responses they were trying out at that moment, along with what they remembered from earlier cycles.

AIM\protect\raisebox{-2pt}{\protect\includegraphics[height=10pt]{figures/AIM-icon.pdf}}'s actions cover a single cycle, from entry through governance. Participants described few practices aimed at retention (\Cref{subsec:result-RQ1}), so extending AIM\protect\raisebox{-2pt}{\protect\includegraphics[height=10pt]{figures/AIM-icon.pdf}} to that stage, where contributors decide whether to stay and some become mentors, is a natural direction for future work. Because AIM\protect\raisebox{-2pt}{\protect\includegraphics[height=10pt]{figures/AIM-icon.pdf}} took shape during the 2026 cycle, studying it in use falls to a later cycle. How organizations adapt AIM\protect\raisebox{-2pt}{\protect\includegraphics[height=10pt]{figures/AIM-icon.pdf}} to their projects, and what it asks of their mentors, could be studied within a single cycle. Whether AIM\protect\raisebox{-2pt}{\protect\includegraphics[height=10pt]{figures/AIM-icon.pdf}} helps contributors stay and mentors return will take several cycles to see, and those are the outcomes that matter most for the program's sustainability.


\section{Related Work} \label{sec:background}

Our work draws on research on OSS mentoring and sustainability and on GenAI-assisted learning and governance.

\textbf{OSS mentoring and sustainability.} OSS mentoring helps newcomers develop project knowledge, technical skills, and social bonds~\cite{feng2024guiding,feng2025mentoring,steinmacher2021mentor,trainer2017mentor,fagerholm2014onboard}. Much of this happens in code review, where mentors transfer knowledge through iterative instruction, feedback, and suggestions~\cite{rigby2013review,bacchelli2013review,sadowski2018review}. Mentoring in these routine activities often remains implicit~\cite{feng2022implicit}. Programs like GSoC give it a formal structure, but in both settings it relies on mentors voluntarily investing their time and effort~\cite{balali2018}. Mentoring is part of the broader interpersonal and organizational work that sustains OSS communities, including maintenance~\cite{geiger2021labor,coelho2017fail}, moderation~\cite{li2021govern,hsieh2023mod,tourani2017conduct}, and contributor retention~\cite{feng2026retention,fang2009retention,ducheneaut2005social}. Mentoring capacity thus limits how many newcomers a community can support. These studies leave open how GenAI changes the effort that mentoring requires.

\textbf{GenAI-assisted learning and governance.} Studies of CodeAid and an AI mentor for OSS explore how GenAI can support learning and onboarding~\cite{kazemitabaar2023codeaid,tan2025aimentor}. A controlled experiment found that code generation improved novices' code-authoring performance without reducing their manual code-modification performance~\cite{kazemitabaar2023codegen}. However, successful task completion can coexist with diminished learning~\cite{shen2026skillformation,bastani2025harm}, weaker code understanding~\cite{martin2026understand,shihab2025understand,prather2023understand}, and persistent metacognitive difficulties~\cite{prather2024metacog}, so completed work may say less about what its author can do independently. OSS communities have begun to respond. Research shows that AI-generated submissions strain community capacity~\cite{afroz2026slop}, a socio-technical roadmap identifies opportunities and risks for OSS learning and governance~\cite{feng2026charting}, and policy analyses characterize boundaries on GenAI use, disclosure, and human involvement~\cite{yang2026governance,hora2026policies}. These studies focus on learners, tools, and project policies. They say little about how mentors in a formal program assess newcomers and apply evolving GenAI rules, which is the focus of our study.

\section{Conclusion} \label{sec:conclusion}

GenAI has made it easy for a newcomer's work to look finished, and harder for a mentor to know what the newcomer has learned. In GSoC, mentors are making up the difference with their own time, the same time that turns this year's contributors into next year's mentors. AIM\protect\raisebox{-2pt}{\protect\includegraphics[height=10pt]{figures/AIM-icon.pdf}} is our attempt to help organizations choose responses their mentors can sustain, 
so that checking work can again be a way of mentoring. As P2 put it, \resultsquote{One principle of GSoC is that we want people to learn... [not] to be completely dependent upon AI... [but] we also want them to use AI because AI is going to be in trend... we don't want them to get backtracked.} If programs like GSoC can protect both their newcomers' learning and their mentors' energy, GenAI may yet help grow the next generation of OSS mentors instead of wearing out the current one.

\section*{Data Availability}
Our interactive dashboard \cite{suppl-web} hosts research artifacts, including the interview guide, screening survey, and qualitative codebooks.

\begingroup
\footnotesize
\bibliographystyle{IEEEtran}
\bibliography{references}
\endgroup


\end{document}